\documentclass[showpacs, aps, prb, unsortedaddress]{revtex4-1}

\usepackage{amssymb}
\usepackage{latexsym}
\usepackage{amsmath}
\usepackage{graphicx}

\begin{document}

\title{Evanescent Wave Boundary Layers in Metamaterials and Sidestepping them through a Variational Approach}

\date{\today}
\author{Ankit Srivastava}
\thanks{Corresponding Author}
\affiliation{Department of Mechanical, Materials, and Aerospace Engineering,
Illinois Institute of Technology, Chicago, IL, 60616
USA}
\email{asriva13@iit.edu}
\author{John R. Willis}
\affiliation{Department of Applied Mathematics and Theoretical Physics, University of Cambridge, Wilberforce Road, Cambridge, CB3 0WA}

\begin{abstract}
All metamaterial applications are based upon the idea that extreme material properties can be achieved through appropriate dynamic homogenization of composites. This homogenization is almost always done for infinite domains and the results are then applied to finite samples. This process ignores the evanescent waves which appear at the boundaries of such finite samples. In this paper we first clarify the emergence and purpose of these evanescent waves in a model problem consisting of an interface between a layered composite and a homogeneous medium. We show that these evanescent waves form boundary layers on either side of the interface beyond which the composite can be represented by appropriate infinite domain homogenized relations. We show that if one ignores the boundary layers then the displacement and stress fields are discontinuous across the interface. Therefore, the scattering coefficients at such an interface cannot be determined through the conventional continuity conditions involving only propagating modes. Here we propose an approximate variational approach for sidestepping these boundary layers. The aim is to determine the scattering coefficients without the knowledge of evanescent modes. Through various numerical examples we show that our technique gives very good estimates of the actual scattering coefficients beyond the long wavelength limit.

\end{abstract}

\maketitle

\section{Introduction}
Metamaterials are artificially designed composite materials which can exhibit properties that are not found in naturally occuring materials. These properties can be electromagnetic \cite{fleury2014cloaking,simovski2009material}, acoustic \cite{chen2010acoustic,craster2012acoustic,kadic2013metamaterials,norris2015acoustic}, or elastodynamic \cite{srivastava2015elastic}. In the context of electromagnetism these properties refer to magnetic permeability and electrical permittivity. For acoustic metamaterials they refer to bulk modulus and density and for elastodynamic metamaterials they refer to moduli (bulk, shear, anisotropic) and density. Irrespective of the different properties which metamaterials research in different fields target, their final aim is the same. Metamaterials research seeks to design composite materials for the fine-tuned, predominantly frequency dependent control of the trajectory and dissipation characteristics of the applicable waves.

Metamaterial properties are generally achieved through an appropriate dynamic homogenization technique which relates a microstructure to its frequency dependent homogenized properties. Currently there are two main ways of doing this. The first is based upon asymptotic methods \cite{bensoussan1978asymptotic,sanchez1980non,bakhvalov1989homogenisation,parnell2006dynamic,andrianov2008higher,craster2009mechanism,craster2010high,antonakakis2013asymptotics,antonakakis2014homogenisation} and the second is based upon field averaging methods \cite{willis1981avariational,willis1981bvariational,willis1983overall,willis1984variational,willis2009exact,nemat2011overall,shuvalov2011effective,willis2011effective,srivastava2012overall,norris2012analytical,willis2012construction}. In addition to these there are scattering measurements based analytical and experimental techniques as well. While simple in principle these run the risk of resulting in properties which violate basic thermodynamic laws \cite{simovski2009material,srivastava2015causality}. Within the established routes of dynamic homogenization the process is the following: periodic boundary conditions are assumed over a unit cell which gives rise to wave solutions of the Bloch form. The fields resulting from these Bloch waves are then homogenized which gives rise to frequency dependent effective properties for the composite metamaterial.

Dynamic homogenization serves as a route for realizing the challenging properties required by the application areas of metamaterials research (transformation acoustics \cite{norris2008acoustic}, elastodynamics \cite{milton2006cloaking,norris2011elastic} etc.) The assumption is that the regions which require a certain set of metamaterial properties can be realized through a composite whose dynamically homogenized properties are the same as the desired properties. A deeper assumption here is that the homogenized properties which were initially calculated for infinite domains can now be applied to non-infinite domains. This assumption is not always correct. The free space homogenized properties may or may not apply to non-infinite cases and this has been explicitly shown to be the case by various researchers \cite{srivastava2014limit,willis2013some,joseph2015reflection}. The reason for the failure of this assumption is subtle. Since dynamically homogenized properties are calculated for free space waves, they do not allow for any evanescent modes. However, the composites which are supposed to realise these properties in a finite setting support evanescent modes. When such composites are interfaced with other regions then these evanescent modes are integral in satisfying displacement and stress continuity across the interface. Without these evanescent modes, however, the stress and displacement fields are discontinuous. Therefore, if a complete correspondence is to be maintained between the composite and the dynamically homogenized region which it is supposed to represent (in the sense that scattering from the two should be equivalent) then the relations between the displacement and stress fields at the interface between the homogenized region and its surrounding region are not ones of simple continuity. In other words, the interface conditions are indeterminate and the true scattering coefficients cannot be determined through simple displacement and stress continuity relations. This is clearly a fundamental issue which results from ignoring evanescent modes in the process of dynamic homogenization.

This issue has been recognised in the electromagnetics community for several decades. There have, therefore, been efforts to account for the effect of these evanescent waves. The predominant technique for doing so is through the inclusion of the so called Drude transition layers \cite{drude1925theory,strachan1933reflexion,simovski2009material}. A transition layer is an artificial layer which is placed between two materials (a metamaterial and a homogeneous material for instance) and which has suitably chosen material properties. The choice is geared towards producing the scattering coefficients which would have resulted had the evanescent modes not been ignored in the original problem. The concept of introducing a transition layer with simple physical properties to simulate the effect of the actual boundary layers created by evanescent waves is an appealing one but defining how to obtain a suitable set of parameters is not straightforward except in the ``quasistatic'', or ``homogenization'' range of frequencies, for which it was first introduced. Furthermore, it is not clear how these transition layers can be made to handle cases where there are multiple propagating modes.

In this paper we propose a different approach for taking into account the effect of the evanescent waves. The approach is illuminated through its application to a model problem which has been considered in recent papers \cite{willis2015negative,nemat2015anti,srivastava2016metamaterial}. It is based upon the observation that at an interface, the energy flux balance only contains contributions from the propagating modes. We, therefore, propose to determine the scattering coefficients of the propagating modes by insisting that they exactly satisfy the energy flux balance while minimizing the displacement discrepancy across the interface. Thus the energy flux is exactly satisfied but the continuity conditions are only approximately satisfied. We show that the process works very well in estimating the ``exact'' scattering coefficients for a wide range of cases including those which are far beyond the long wavelength limit. The process is also general enough to potentially apply to more complex 2- and 3-D cases.

\section{Bloch Waves in the Laminate}

Following \cite{willis2015negative} we define our laminate as a periodically layered structure in the $x_1$ direction with the layer interfaces in the $x_2$--$x_3$ plane and infinite in this plane. In the direction of periodicity the laminated composite is characterized by a unit cell $\Omega$ of length $h$ ($0\leq x_1\leq h$). For our purposes the unit cell is composed of two material layers with shear moduli $\mu_1,\mu_2$, densities $\rho_1,\rho_2$, and thicknesses $h_1,h_2$ respectively. If anti-plane shear waves are propagating in the laminate then the only nonzero component of displacement is taken to be $u_3$ which has the functional form $u_3(x_1,x_2,t)$. Within the $i^{\rm th}$ layer ($i=1,2$) it satisfies the following equation of motion:
\begin{eqnarray}
\label{eAntiPlane}
\displaystyle u_{3,11}+u_{3,22}=\frac{1}{c^2_i}\ddot{u}_3
\end{eqnarray}
where $c_i=\sqrt{\mu_i/\rho_i}$. The displacement gives rise to stress fields $\sigma_{13}(x_1,x_2,t),\sigma_{23}(x_1,x_2,t)$. The shear stress component $\sigma_{13}$ and displacement $u_3$ are continuous at the material interfaces. Across an interface between layers $i$ and $i+1$ at $x_1=x^i$:
\begin{equation}
\label{eContinuity}
\mathbf{v}^i|_{x_1=x_i}\equiv\begin{pmatrix}\sigma_{13}(x^i,x_2,t) \\ u_3(x^i,x_2,t) \end{pmatrix}^i=\mathbf{v}^{i+1}|_{x_1=x_i}
\end{equation}
Due to the periodicity of the laminate, the displacement and stress fields follow Bloch-periodicity conditions. Generally we have $\mathbf{v}\equiv\tilde{\mathbf{v}}(x_1)e^{i(\omega t-K_1x_1-k_2x_2)}$, where $\tilde{\mathbf{v}}(x_1)$ is periodic with $\Omega$; specifically, for the displacement field, we have:
\begin{eqnarray}
\label{eBlochAntiPlane}
\displaystyle u_3(x_1,x_2,t)=\tilde{u}(x_1)\mathrm{e}^{i(\omega t-K_1x_1-k_2x_2)}
\end{eqnarray}
The wavenumber component $k_2$ must be continuous across the layers to satisfy Snell's law. The other nonzero stress component $\sigma_{23}$ has a similar Bloch-periodic form but $\tilde\sigma_{23}(x_1)=-ik_2\mu(x_1)\tilde u(x_1)$ is not continuous across material interfaces. 
By using the general solutions to the governing equation (\ref{eAntiPlane}), the continuity of traction and displacement at the interfaces (\ref{eContinuity}), and the Bloch formulation (\ref{eBlochAntiPlane}), we can formulate a Transfer Matrix formulation ($x_2,\omega$ dependence suppressed):
\begin{equation}
\label{eTMM}
\mathbf{v}(h)=M\mathbf{v}(0)=\lambda\mathbf{v}(0)
\end{equation}
where the eigenvalue $\lambda=e^{-iK_1h}$. Quantities in the above equation depend upon assumed values of $\omega,k_2$. The solutions to the eigenvalue problem above furnish the wavenumber $K_1$ and the modeshape for which (\ref{eBlochAntiPlane}) satisfies the governing equation. The wavenumber solutions themselves come from the following equation:
\begin{equation}
\label{eAntiPlaneS}
\cos(K_1h)=\frac{1}{2}\mathrm{tr}(M)
\end{equation}
so that if $K_1$ is a solution then so are $\pm(K_1\pm 2n\pi/h)$ for all integer $n$.
Consider two different solutions of the current problem (details in \cite{willis2015negative}):
\begin{eqnarray}
\displaystyle u_3(x_1,x_2,t)=\tilde{u}(x_1)\mathrm{e}^{i(\omega t-K_1x_1-k_2x_2)}\\
\displaystyle v_3(x_1,x_2,t)=\tilde{v}(x_1)\mathrm{e}^{i(\omega t-K_1x_1-\bar{k}_2x_2)}
\end{eqnarray}
It can be shown that as long as $\omega^2,K_1,k_2^2,\bar{k}_2^2$ are real and $k_2^2\neq \bar{k}_2^2$, the modeshapes $\tilde{u},\tilde{v}$ are orthogonal with respect to the weight $\mu$:
\begin{eqnarray}
\label{ortho}
\langle\tilde{u},\mu\tilde{v}\rangle=\frac{1}{h}\int_0^h\tilde{u}\mu\tilde{v}^*\mathrm{d}x_1=0
\end{eqnarray}
Throughout the sequel, the modeshapes are normalized so that $\langle \tilde{u},\mu\tilde{u}\rangle=\bar\mu$, where $\bar{\mu}$ is the mean modulus of the laminate.  

\section{Normal Mode Decomposition}

\begin{figure}[htp]
\centering
\includegraphics[scale=.5]{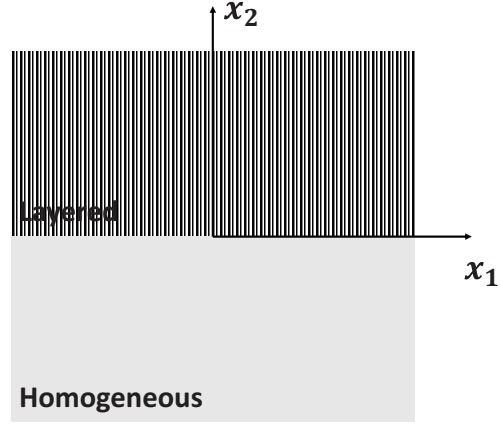}
\caption{Schematic of the interface problem.}\label{fschematic}
\end{figure}
We now consider an interface between a homogeneous medium with shear modulus $\mu_0$ and the layered composite. The interface itself can be placed at any angle with the layers but presently we assume that it is along $x_2=0$ (Fig. \ref{fschematic}). The layered medium is in the region $x_2>0$ with the layers being parallel to the $x_2$ axis. A plane harmonic wave is incident at the interface from the homogeneous medium. This wave sets up an infinite number of transmitted and an infinite number of reflected waves. A finite number of these are propagating waves and the rest are evanescent waves. The incident, transmitted, and reflected fields are written down as:
\begin{eqnarray}
&\hbox{Incident Field:}&\quad A\exp\left[i(\omega t-k\sin\theta x_1-k\cos\theta x_2)\right]\\
&\hbox{Transmitted Field:}&\quad \sum_{m=0}^\infty T_m\tilde{u}_m(x_1)\exp\left[i(\omega t-k\sin\theta x_1-k_2^{(m)} x_2)\right]\\
&\hbox{Reflected Field:}&\quad \sum_{n=-\infty}^\infty R_nU_n(x_1)\exp\left[i(\omega t-k\sin\theta x_1+\kappa^{(n)} x_2)\right]
\end{eqnarray}
where $U_n(x_1)=e^{-i2n\pi x_1/h}$ and $\kappa^{(n)}\equiv\left[k^2-(k\sin(\theta)+2n\pi/h)^2\right]^{1/2}$ is taken either as positive real or negative imaginary to prevent exponential rise in $x_2<0$. The wavenumber components $k_2^{(m)}$ are either positive real or negative imaginary and satisfy (\ref{eAntiPlaneS}) with $K_1 = k\sin\theta$. At any given frequency, the transmitted field will consist of $M$ propagating modes and infinitely many evanescent modes in the $x_2$ direction. The problem of determining the scattering parameters can be solved to any required degree of precision by considering a sufficiently large number of terms in the normal mode expansions. To facilitate calculations we can restrict the reflected modes to a range of $-N\leq n\leq N$ and transmitted modes to a range of $0\leq m\leq 2N$ such that $2N+1>M$. This allows us to consider all propagating transmitted modes in the expansion. With this, the displacement ($u_3$) and stress ($\sigma_{23}$) continuity are given by (exponential terms suppressed):
\begin{eqnarray}\label{bc}
\displaystyle \sum_{m=0}^{2N}\bar{T}_m\tilde{u}_m(x_1)-\sum_{n=-N}^{N}\bar{R}_nU_n(x_1)\approx 1\\
\displaystyle \mu(x_1)\sum_{m=0}^{2N}k_2^{(m)}\bar{T}_m\tilde{u}_m(x_1)+\mu_0\sum_{n=-N}^{N}\kappa^{(n)}\bar{R}_nU_n(x_1)\approx \mu_0k\cos\theta
\end{eqnarray}
where $\bar{R}_n=R_n/A,\bar{T}_n=T_n/A$. Note that strict equality only holds in the limit of $N\rightarrow\infty$. 
The above can be transformed into a system of $2(2N+1)$ equations in as many variables through the application of the orthogonality condition (\ref{ortho}). Specifically we have
\begin{eqnarray}
\label{scattering}
\displaystyle 
\begin{bmatrix}
    [M_1] & [M_2]\\
    [M_3] & [M_4]
\end{bmatrix}\{S\}=\{I\}
\end{eqnarray}
where $S$ is a column vector of size $2(2N+1)$ with elements $\bar{T}_0,...\bar{T}_{2N},\bar{R}_{-N},...\bar{R}_{N}$. Submatrices $[M_i]$ are square matrices of sizes $(2N+1)\times(2N+1)$ with the following nonzero elements:
\begin{eqnarray}
\displaystyle [M_1]_{ij}=\delta_{ij},\quad [M_2]_{ij}=-\langle U_{j-N},\mu\tilde{u}_i\rangle/{\bar\mu}\\
\displaystyle [M_3]_{ij}=k_2^{(i)}\delta_{ij},\quad [M_4]_{ij}=\kappa^{(j-N)}\langle U_{j-N},\mu_0\tilde{u}_i\rangle/{\bar\mu}\\
i,j=0,...2N
\end{eqnarray}
and $I$ is a column vector of size $2(2N+1)$ with elements
\begin{eqnarray}
\displaystyle I_{i}=\langle 1,\mu\tilde{u}_i\rangle/{\bar\mu},\quad 0\leq i\leq 2N\\
\displaystyle =k\cos\theta\langle 1,\mu_0\tilde{u}_i\rangle/{\bar\mu},\quad i>2N
\end{eqnarray}
We can further write down an energy flux balance based on the scattering coefficients::
\begin{eqnarray}
\label{energyidentity}
E=\frac{{2}}{\mu_0\omega k\cos\theta_i}\sum_{i=1}^{2(2N+1)}|\bar{S}_i|^2 \langle\mathcal{P}_2\rangle^{(i)}=1
\end{eqnarray}
where $\langle\mathcal{P}_2\rangle$ is the 2-component of the time and unit cell averaged Poynting vector for the $i^{\rm th}$ mode.

\section{Evanescent Field as a Boundary Layer}

\begin{figure}[htp]
\centering
\includegraphics[scale=.7]{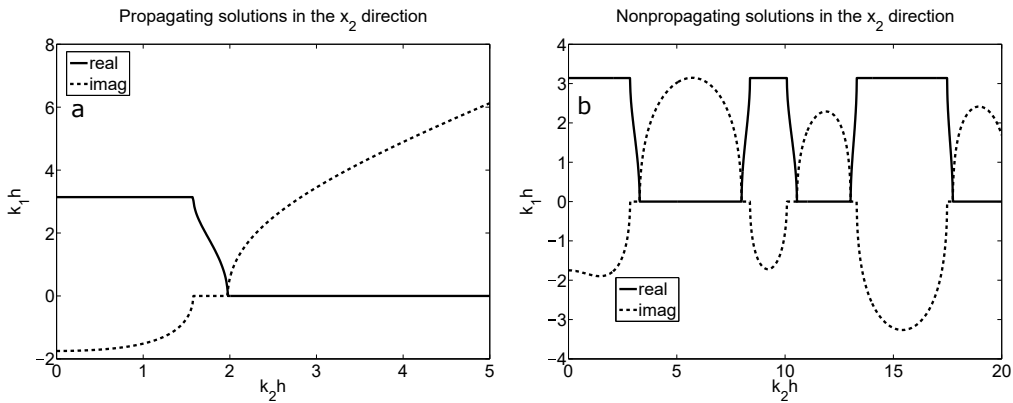}
\caption{$k_1h$--$k_2h$ plots for the laminate at 200 kHz.}\label{f200}
\end{figure}

To illustrate the role of evanescent waves we consider the general example that was treated in \cite{willis2015negative}. The homogeneous medium is taken to be Aluminum ($\mu_0=26 \mathrm{GPa}$, $\rho_0=2700 \mathrm{kg/m}^3$) and the laminated composite is composed of two materials of thicknesses $h_1=0.003$m (Epoxy: $\rho_1=1180$ kg/m$^3$, $\mu_1=3$ GPa) and $h_2=0.0013$m (Steel: $\rho_2=8000$ kg/m$^3$, $\mu_2=80$ GPa). We take the frequency of excitation to be 200 kHz in which case the first propagating band is fully developed (Fig. \ref{f200}). 
At this frequency there is one fully propagating mode in the laminate and infinitely many modes which are propagating in the $x_1$ direction and non-propagating in the $x_2$ direction. For the interface configuration under consideration it is the set of non-propagating modes which forms the evanescent boundary layer adjacent to the interface and on the side of the laminate. There is another evanescent boundary layer formed on the side of the homogeneous material which is determined by those reflected modes for which $\kappa^{(n)}$ is negative and imaginary. Concentrating for the present on the laminate side of the interface, evanescent modes in the $x_2$ direction have the functional dependence $\exp(-|k_2^{(m)}|x_2)$ where $k_2^{(m)}$ is negative-imaginary. Due to exponential decay, the influence of those modes with large $k_2^{(m)}$ components is  smaller than those with smaller components (assuming comparable transmission coefficients). At 200 kHz and for $k_1h=1.5$ the first evanescent mode has $|k_2^{(2)}|h=3.096$ ($k_2^{(1)}$ defines the propagating mode). At $x_2=1.4$mm from the interface ($|k_2^{(2)}|x_2=1$) the amplitude of this mode is reduced to $37\%$ of its value at the interface. At this location the second and third evanescent modes have $7\%$ and $3.5\%$ of their respective amplitudes at the interface. Unless there is significant disparity in the transmission coefficients of the evanescent modes, it is safe to say that if the first evanescent mode may be neglected beyond a certain distance from the interface then all higher evanescent modes may also be neglected beyond this distance. We term this distance the boundary layer thickness $t_e$, defined (for the purpose of this discussion) as the distance from the interface at which the evanescent wave with the smallest magnitude of its wavenumber reaches $10\%$ of its amplitude at the interface. For the present case when $|k_2^{(2)}|h=3.096$, we have $t_e=0.7437h$. At locations which are more than $t_e$ away from the interface the scattered field can be taken to consist only of the propagating solutions. On the homogeneous material side, the boundary layer thickness is similarly determined by that imaginary $\kappa^{(n)}$ which has the smallest magnitude. The boundary layer thickness changes with both the angle of incidence and the frequency. It can be arbitrarily large at those frequencies and incidence angle combinations where the first evanescent mode has a vanishingly small imaginary part of the wavenumber. In metamaterial applications one generally seeks to replace finite composite samples with their free space homogenized constitutive properties. Inherent in this process is the assumption that the evanescent modes which are invariably generated at an interface can be neglected away from it. It is clear from the above that even at low frequencies there may exist cases where no such replacement is possible because the influence of the evanescent modes may persist throughout the sample due to a large value of $t_e$.

\subsection{Continuity Conditions and Energy Conservation}

While the evanescent modes are required for the satisfaction of the boundary conditions (\ref{bc}), they do not enter the energy conservation equation (\ref{energyidentity}). Conservation of energy must, therefore, emerge from the satisfaction of boundary conditions. To show that this is indeed true we split the energy conservation equation into the $m_t$ transmitted and $n_r$ reflected propagating components (since nonpropagating components do not contribute):
\begin{eqnarray}\label{energy2}
\tilde{\mu}\sum_{\tilde{k}_2^{(m)}\;{\rm real}}|\bar{T}_m|^2 \tilde{k}_2^{(m)}+\sum_{\tilde\kappa^{(n)}\;{\rm real}}|\bar{R}_n|^2 \tilde{\kappa}^{(n)}=1
\end{eqnarray}
where $\tilde{\mu}=\bar{\mu}/\mu_0,\tilde{k}_2^{(m)}={k}_2^{(m)}/k\cos\theta, \tilde{\kappa}^{(n)}={\kappa}^{(n)}/k\cos\theta$. Rearranging the boundary conditions:
\begin{eqnarray}\label{bc2}
\displaystyle \nonumber \sum_{m=0}^{\infty}\bar{T}_m\tilde{u}_m(x_1)=1+\sum_{n=-\infty}^{\infty}\bar{R}_nU_n(x_1)\\
\displaystyle \frac{\mu(x_1)}{\mu_0}\sum_{m=0}^{\infty}\tilde{k}_2^{(m)}\bar{T}_m\tilde{u}_m(x_1)=1-\sum_{n=-\infty}^{\infty}\tilde{\kappa}^{(n)}\bar{R}_nU_n(x_1)
\end{eqnarray}
Note that we have extended the summation to infinity as strict equality only holds in the limit of considering all available modes. Taking the complex conjugate of the displacement boundary condition, multiplying respective sides with the stress boundary condition, and averaging over the unit cell gives:
\begin{eqnarray}
\displaystyle \nonumber \frac{1}{\mu_0}\sum_{m=0}^{\infty}\sum_{m'=0}^{\infty}\tilde{k}_2^{(m)}\bar{T}_m\bar{T}_{m'}^*\langle\tilde{u}_m,\mu\tilde{u}_{m'}\rangle=1-\sum_{n=-\infty}^{\infty}\sum_{n'=-\infty}^{\infty}\tilde{\kappa}^{(n)}\bar{R}_n\bar{R}_{n'}^*\langle U_n,U_{n'}\rangle\\
+\sum_{n'=-\infty}^\infty\bar{R}_{n'}^*\langle 1,U_{n'}\rangle-\sum_{n=-\infty}^\infty\tilde{\kappa}^{(n)}\bar{R}_{n}\langle U_n,1\rangle
\end{eqnarray}
Now we note that $\langle U_n,U_{n'}\rangle
=\delta_{nn'}$, $\langle\tilde{u}_m,\mu\tilde{u}_{m'}\rangle=\bar\mu\delta_{mm'}$, and $U_0=\tilde{\kappa}^{(0)}=1$. Considering these and taking the real part of the above equation results directly in the energy conservation equation (\ref{energy2}). Note that taking the real part automatically constrains the equality to only the propagating modes since the contribution from the evanescent modes is strictly imaginary. Another equality may be obtained by taking the imaginary part of the above equation:
\begin{eqnarray}\label{energy3}
\tilde{\mu}\sum_{m}|\bar{T}_m|^2 |\tilde{k}_2^{(m)}|+\sum_{n}|\bar{R}_n|^2 |\tilde{\kappa}^{(n)}|=2\mathcal{I}(\bar{R}_0)
\end{eqnarray}
where the summations now only include all the transmitted and reflected evanescent modes. The satisfaction of the boundary conditions (\ref{bc2}), therefore, automatically implies the satisfaction of energy conservation on propagating modes (\ref{energy2}) and an additional conservation relation on the 
evanescent modes (\ref{energy3}). Eq. (\ref{energy3}) has the interesting consequence of bounding the amplitudes of the evanescent modes by the imaginary part of the reflected propagating mode of order 0.

\subsection{Boundary Layers and the Role of Evanescent Waves in Satisfying Continuity Conditions}

Within the boundary layers the relevant stress and displacement components vary continuously with $x_2$, for any fixed $x_1$, between the two values which correspond to the free space propagating waves on both the reflected and transmitted sides. To show this we use the modified form of the interface conditions (\ref{bc2}) and term the left and right sides of the displacement equation $u_t$ and $u_r$ respectively. Similarly $\sigma_t$ and $\sigma_r$ refer to the analogous $\sigma_{23}$  stress components.
\begin{figure}[htp]
\centering
\includegraphics[scale=.7]{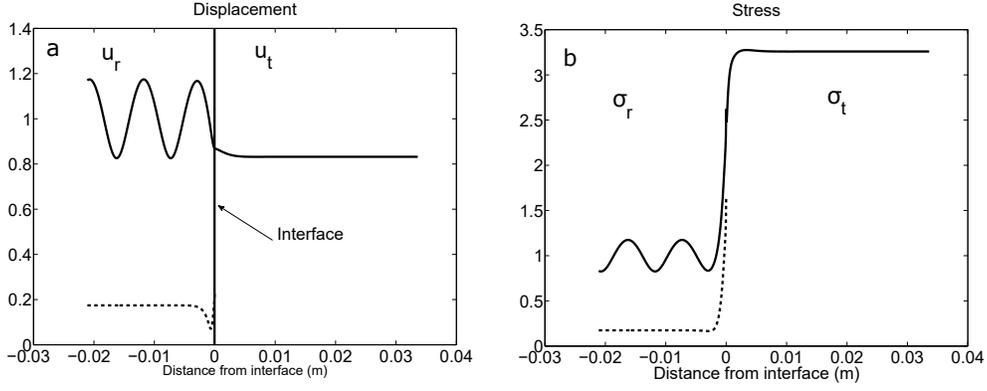}
\caption{Displacement and stress profiles across the interface.}
\label{f200aI}
\end{figure}
Fig. \ref{f200aI} shows the variation of the displacement and stress fields (absolute values) as a function of the distance from the interface. Positive values of distances are into the laminate and all calculations are done over a line $x_1 = $ constant which bisects any one of the steel laminae, material 2. We have used an angle of incidence of $30^0$ and for this case the boundary layer thicknesses in the laminate and the homogeneous medium are 0.0084m and 0.0053m respectively. To clarify the effect of the boundary layer we have also plotted the absolute values of only the scattered components in the homogeneous medium (dashed curves.) Since this case only has one propagating transmitted mode and one propagating reflected mode, the absolute values of the scattered field beyond the boundary layer are constants. This is evident from the dashed curves on the homogeneous side and the solid curves on the laminate side wherein the absolute values of the scattered field stabilize to constants beyond the boundary layer. The sinusoidal variations of the absolute values of $u_r,\sigma_r$ are due to the combination of two waves with opposing wavenumbers in the $x_2$ directions. At the interface the purpose of the evanescent waves is to match $u_r$ and $\sigma_r$ with $u_t$ and $\sigma_t$ respectively. We define $[u]=u_r-u_t$ and $[\sigma]=\sigma_r-\sigma_t$ as measures of how well the continuity conditions are satisfied. These are functions of $x_1$ and are also dependent upon the magnitudes of the displacement and stress terms. To understand how well continuity conditions are being satisfied we average and normalize these measures as $\bar{u}=\langle[u],[u]\rangle/\langle u_r,u_r\rangle$ and $\bar{\sigma}=\langle[\sigma],[\sigma]\rangle/\langle \sigma_r,\sigma_r\rangle$.
\begin{figure}[htp]
\centering
\includegraphics[scale=.55]{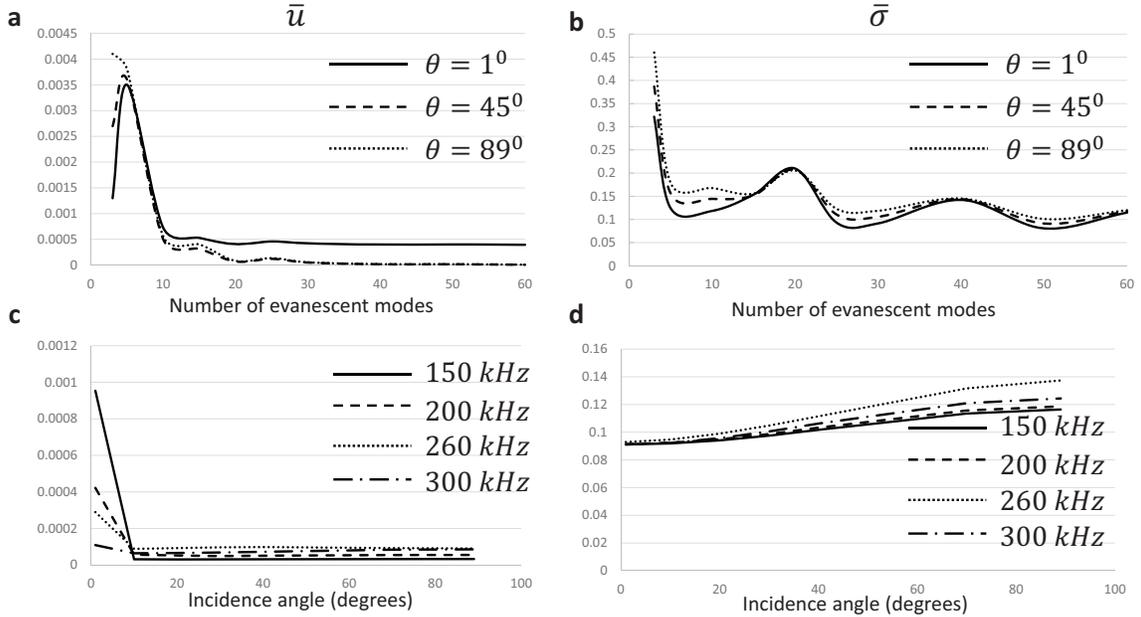}
\caption{$a,b: \bar{u},\bar{\sigma}$ as a function of the number of evanescent modes, c,d: $\bar{u},\bar{\sigma}$ as a function of the incidence angle}\label{fbc}
\end{figure}
Figs. \ref{fbc}a,b plot $\bar{u},\bar{\sigma}$ as functions of the number of evanescent modes. It shows that the displacement continuity condition is satisfied better than the stress continuity condition for a given number of evanescent modes in the expansion. For the considered case $\bar{\sigma}/\bar{u}$ is generally greater than 100. Figs. \ref{fbc}c,d show $\bar{u},\bar{\sigma}$ as functions of the incident angle and for four different frequencies. Note that at 300 kHz there are two propagating transmitted modes whereas for all other frequencies there is one transmitted propagating mode. All calculations are carried out for 30 evanescent modes in the expansion. Again, it is clear that $\bar{u}$ is approximated better than $\bar{\sigma}$ for all cases. For all frequencies except 300 kHz it becomes easier to satisfy the displacement continuity at higher angles of incidences (decreasing $\bar{u}$.) On the contrary, it becomes more difficult to satisfy stress continuity with increasing incidence angles (increasing $\bar{\sigma}$.) $\bar{\sigma}$ decreases slowly with additional evanescent modes at all angles. 
\begin{figure}[htp]
\centering
\includegraphics[scale=.65]{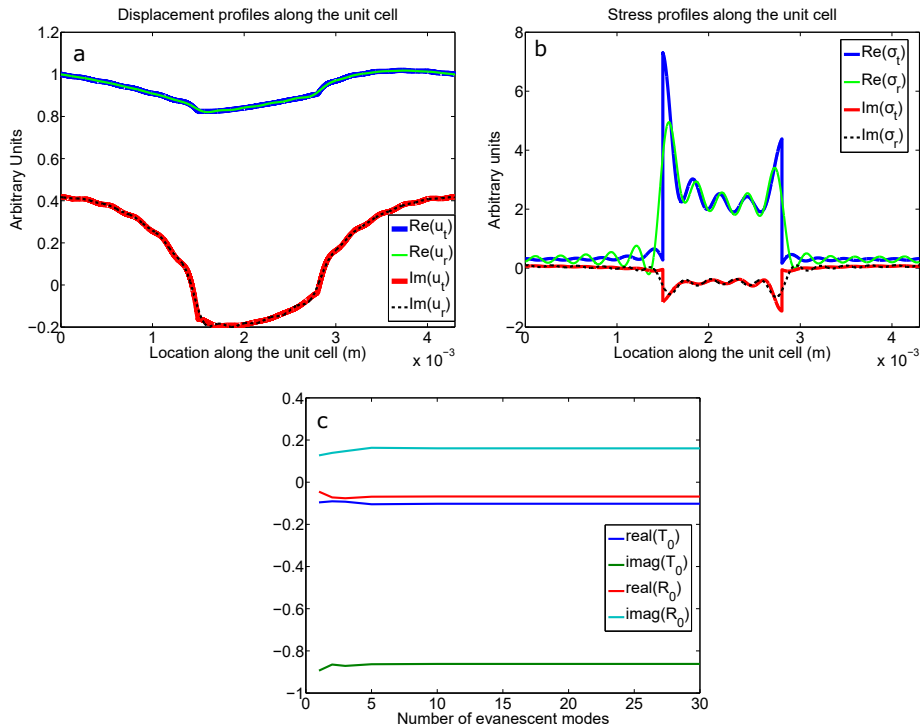}
\caption{a. $u_t,u_r$ along the unit cell, b. $\sigma_t,\sigma_r$ along the unit cell, c. scattered coefficients as a function of the number of evanescent modes in the expansion. Angle of incidence is $30^0$ and frequency is 200 kHz.}\label{fTRE}
\end{figure}
Fig. \ref{fTRE}a,b shows why it is far easier to satisfy displacement continuity than stress continuity. For 200 kHz, $30^0$ incidence, and 30 evanescent modes, we plot $u_r,u_t,\sigma_r,\sigma_t$ as a function of the location along the interface. While $u_t$ is continuous along the unit cell, $\sigma_t$ is discontinuous at the layer transitions. This is not surprising as while the displacement is required to be continuous across the layers, there is no such continuity requirement on $\sigma_{23}$ across the layers which are in the $x_2$--$x_3$ plane. $\sigma_r$ on the other hand will necessarily be continuous as it is composed of trigonomentric functions. To get a good match of $\sigma_r$ with $\sigma_t$, it is clear that evanescent modes with high wavenumbers are required which can adequately approximate the discontinuous jump. On the other hand only relatively few evanescent waves are required to match $u_r$ with $u_t$ which are both necessarily continuous functions. It is further notable that the coefficients for the propagating modes do not change significantly as the number of evanescent modes in the expansion are increased (Fig. \ref{fTRE}c). This behavior is seen for frequency and incidence angle combinations mentioned in Figs. \ref{fbc}c,d. From the above arguments and especially for higher angles of incidence it is clear that the inclusion of a greater number of evanescent modes in the expansion will serve primarily to improve the stress continuity. 

\section{A Variational Approach for Sidestepping the Boundary Layers}

It is clear from the above treatment that beyond the boundary layers the scattered field on both sides of the interface can be safely taken to comprise only of the propagating modes. Therefore, in these zones the material behavior can be described by dynamically homogenized free space effective properties. The issue is the complicating presence of the boundary layers. Given two homogeneous materials and an interface between them the behavior of a wave which impinges upon it is determined by the stress and displacement continuity relations across the interface. However, given a homogenized metamaterial, a homogeneous medium (or another homogenized metamaterial) and an interface between them, how should one determine the behavior of an impinging wave? It is clear from Fig. \ref{f200aI} that stress and displacement are not continuous across the interface if one neglects the boundary layers on either side. The above question can be rephrased into the following: given only the propagating modes of the homogeneous and homogenized media and an interface between them, how can we determine the scattered field resulting from an impinging wave? 

This question is of practical importance in the area of metamaterials research, the primary concern of which is to achieve extreme material properties through appropriate dynamic homogenization techniques. These homogenized properties are almost always calculated for free space propagating waves and then applied to finite or semi-infinite samples. The boundary effects, therefore, are generally neglected. The free space homogenized properties may or may not apply to non-infinite cases and this has been explicitly shown to be the case by various researchers \cite{srivastava2014limit,willis2013some,joseph2015reflection}. In the area of electromagnetism the boundary effect is generally taken into consideration through the inclusion of Drude transition layers \cite{drude1925theory,strachan1933reflexion,simovski2009material}. The problem of assigning to such a layer a suitable set of parameters is not straightforward except in the ``quasistatic'', or ``homogenization'' range of frequencies, for which it was first introduced. Simovski \cite{simovski2009material} has reported progress in identifying layers suitable for higher frequencies, though his work considered only a dipole lattice approximation for electromagnetics. His proposed layers were aimed for use in what he termed the ``metamaterial'' range of frequencies, at which the wavelength in the matrix material  was significantly greater than the Bragg wavelength but the possibility of resonance of the dipoles could occur. He considered only normal incidence and suggested but did not prove that exactly the same parameters might be applicable also for oblique incidence.  It is not clear how the Drude layer concept could be developed to accommodate frequencies at which there are more than one propagating transmitted and/or reflected waves.

We propose to determine the scattering coefficients (allowing only for propagating waves) through an indirect route. First we note that the energy conservation equation (\ref{energy2}) only consists of contributions from propagating modes. Second we note from Figs. (\ref{fbc},\ref{fTRE}) that displacement continuity is satisfied more easily than stress continuity and that higher evanescent modes play a significant role in satisfying stress continuity. With these observations we propose to solve the following minimization problem in search of the appropriate scattering coefficients:
\begin{equation}
\begin{aligned}
& \underset{\bar{\mathbf{S}}^P}{\text{minimize}}
& & \langle[u],\mu[u]\rangle/{\bar\mu} \\
& \text{subject to}
& & \phi(\bar{\mathbf{S}}^P)=0
\end{aligned}
\end{equation}
where $\phi$ is the energy constraint:
\begin{equation}
\phi(\bar{\mathbf{S}}^P)=\tilde{\mu}\sum_{\tilde k_2^{(m)}\;{\rm real}}|\bar{T}_m|^2 \tilde{k}_2^{(m)}+\sum_{\tilde \kappa^{(n)}\;{\rm real}}|\bar{R}_n|^2 \tilde{\kappa}^{(n)}-1
\end{equation}
In the above equations, the superscript $P$ refers to the fact that only propagating modes are being considered. $\bar{\mathbf{S}}^P$, as earlier, refers to the normalized values of the scattering coefficients. The minimization problem gives rise to a system of equations through the use of a Lagrange multiplier. If there are $m_t$ propagating transmitted modes and $n_r$ propagating reflected modes then this system is expressed in a matrix form:
\begin{equation}
\left[\mathbf{M}+\lambda\mathbf{N}\right]\{\bar{\mathbf{S}}^P\}+\{\mathbf{I}\}=0
\end{equation}
where $\mathbf{M},\mathbf{N}$ are square matrices of size $m_t+n_r$, $\mathbf{I}$ is a column vector of length $m_t+n_r$, and $\lambda$ is the Lagrange multiplier. We have:
\begin{eqnarray}
\nonumber \displaystyle 
i,j\leq m_t:\quad M_{ij}=\delta_{ij};\quad I_{i}=-\langle 1,\mu\tilde{u}_i\rangle/{\bar\mu}\\
\nonumber \displaystyle 
i\leq m_t,j>m_t:\quad M_{ij}=-\langle U_{j-m_t},\mu\tilde{u}_i\rangle/{\bar\mu};\quad I_{i}=-\langle 1,\mu\tilde{u}_i\rangle/{\bar\mu}\\
\nonumber \displaystyle 
i>m_t,j\leq m_t:\quad M_{ij}=-\langle \tilde{u}_j, \mu U_{i-m_t}\rangle/{\bar\mu};\quad I_{i}=\langle 1, \mu U_{i-m_t}\rangle/{\bar\mu}\\
\displaystyle 
i,j>m_t:\quad M_{ij}=\langle U_{j-m_t}, \mu U_{i-m_t}\rangle/{\bar\mu};\quad I_{i}=\langle 1, \mu U_{i-m_t}\rangle/{\bar\mu}
\end{eqnarray}
Matrix $\mathbf{N}$ is diagonal with components $N_{ij}=2\langle\mathcal{P}\rangle_2^{(i)}/\mu_0\omega k\cos\theta\delta_{ij}$. With the energy constraint as an additional equation, the above is a system of $m_t+n_r+1$ equations in $m_t+n_r+1$ unknown variables. Being nonlinear this system is solved through established gradient descent algorithms. In the following subsections we compare the scattering coefficients which we calculate from the above minimization process with those which are calculated from Eq. (\ref{scattering}). The former considers only propagating modes whereas the latter considers both propagating and evanescent modes.

\begin{figure}[htp]
\centering
\includegraphics[scale=.7]{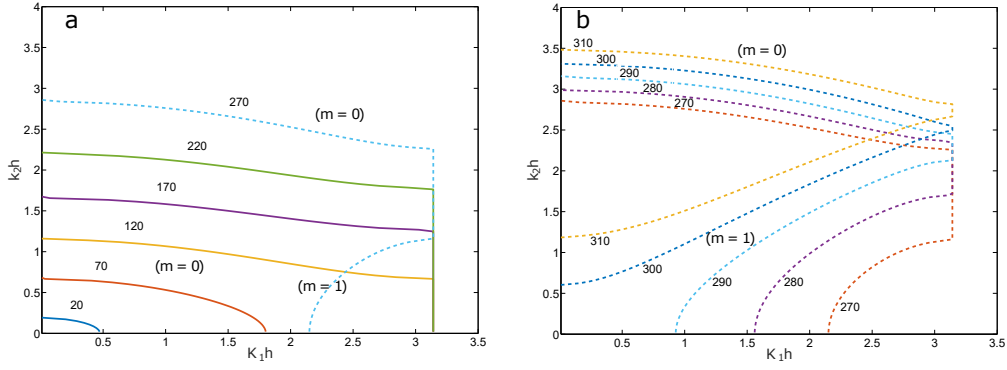}
\caption{Plots of $k_2h$ versus $K_1h$ at several fixed frequencies, between 270 and 310 kHz}\label{ff270-310}
\end{figure}

Before proceeding, it is relevant to refer to Fig. \ref{ff270-310}, which is similar to Fig. 2 of \cite{willis2015negative} but with slightly different ranges of frequency. Fig. \ref{ff270-310}a displays equifrequency contours in the $K_1$--$k_2$ plane in the lower range of frequencies. Except for $f=270$ kHz there is only one propagating mode, to which we assign the label $m=0$. These waves undergo positive refraction (both components of group velocity are positive). At $f=270$ kHz however, there are two transmitted modes for $K_1h$ greater than about 2.2. For incidence from an aluminum half-space, this corresponds to an angle of incidence of approximately 69.4$^\circ$. This additional mode is negatively refracted and is assigned the label $m=1$. It appears first (at $K_1h=\pi$, $k_2h = 0$) at $f\approx 261332$ Hz. Fig \ref{ff270-310}b displays a range of higher frequencies in which there are two propagating modes, $m=0$, positively refracted and $m=1$, negatively refracted. The range of $K_1h$ values over which the mode $m=1$ exists increases as the frequency increases.
The figure does not show it but the frequency above which there are two propagating modes for all $K_1h \in [0,\pi)$ is 296630 Hz.
\subsection{Examples with positive refraction}

\begin{figure}[htp]
\centering
\includegraphics[scale=.6]{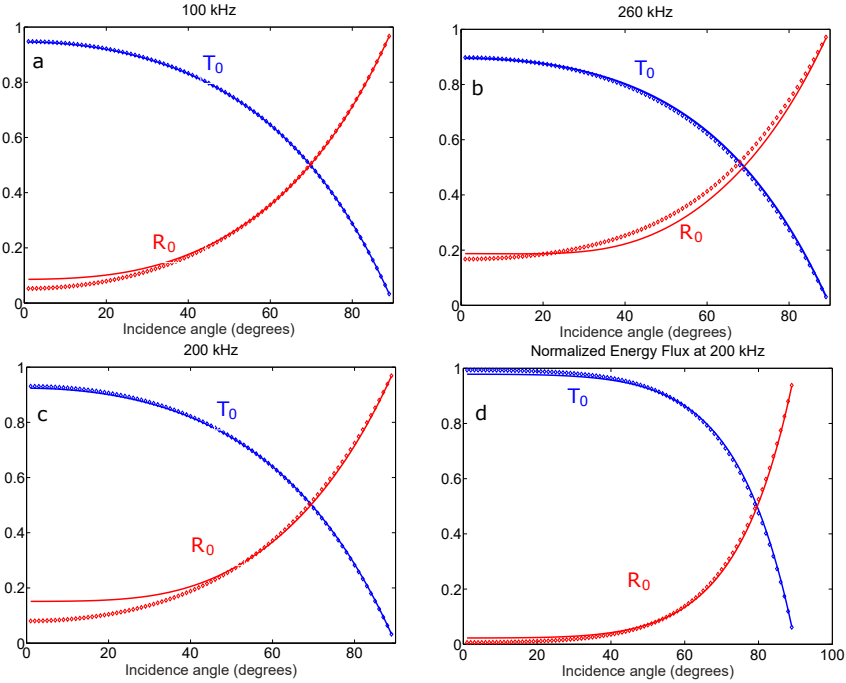}
\caption{Comparison of the absolute values of the calculated scattering coefficients. Solid lines: full calculation; diamonds: ``optimised'' calculation. The fourth plot shows comparisons of the calculated energy fluxes at frequency $f=200$ kHz.}\label{fsingleMode}
\end{figure}

As the first example we compare the scattering coefficients calculated from the two approaches at 100 kHz, 200 kHz and 260 kHz. At these frequencies and for all angles of incidence less than $90^0$, there exist one transmitted propagating mode ($T_0$ mode) and one reflected propagating mode ($R_0$ mode). Fig. \ref{fsingleMode} shows the above mentioned comparison for the absolute values of the scattering coefficients as functions of the angles of incidence. In this and later figures, the values calculated by allowance for 30 modes are plotted as solid lines, while those calculated from the optimization scheme are shown as diamonds. Note that there is no reason why the phase information (real and complex parts) of the scattering coefficients should match for the two approaches. This is due to the fact that the imposed constraint is on energy which depends only upon the magnitudes of the scattering coefficients. The definition of the phase for any wave in the laminate is arbitrary, in any case: if $\tilde u$ is any mode, then $e^{i\theta}\tilde u$ is equally acceptable, for any $\theta$. The good agreement is not surprising in the case of the lowest frequency because 100 kHz is not far beyond what may be regarded as the ``homogenization'' range, in which the evanescent modes contribute little, the transmission and reflection coefficients are real and the energy balance equation provides a legitimate substitute for the equation giving continuity of traction. It can be seen from the figure that optimized results are also close to the ``exact'' results at the two higher frequencies, especially for incidence directions away from normal. They are, in addition, good enough to be useful, even close to normal incidence: the reflection coefficient is small and the reflected energy is proportional to the square of its magnitude. This is demonstrated in Fig. \ref{fsingleMode}d, which shows plots of the approximate and ``exact'' energy fluxes (the individual terms in (\ref{energy2})), for $f=200$ kHz. Although the relative error may be large, the absolute error is small. Thus, the scattered energies for both propagating modes are well estimated without considering the boundary layers by the variational scheme, at all angles of incidence.

\subsection{Examples with negative refraction}

\begin{figure}[htp]
\centering
\includegraphics[scale=.6]{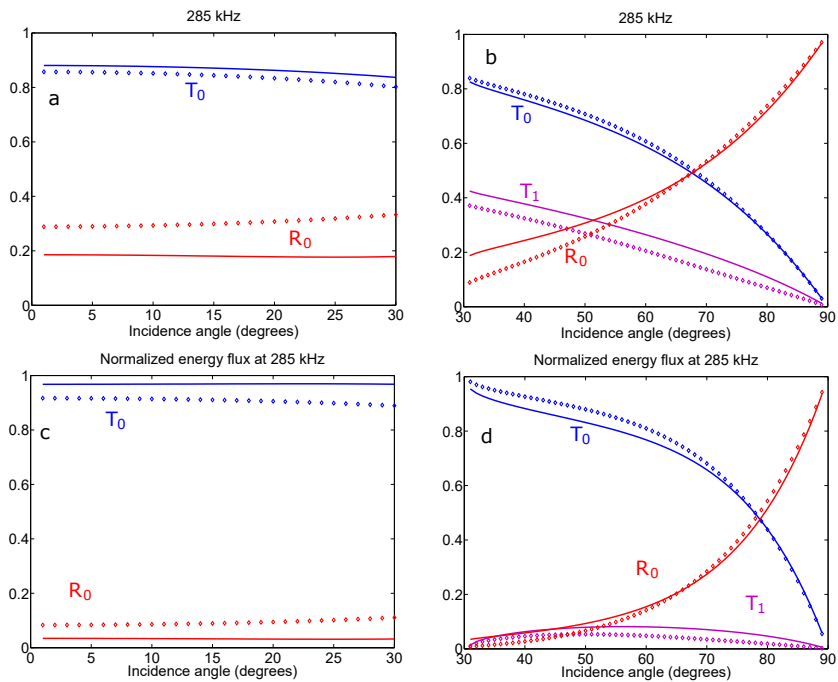}
\caption{Comparison of the absolute values of the calculated scattering coefficients at 285 kHz. For angles of incidence below about 30$^\circ$ there is only one transmitted mode while for higher angles of incidence there are two transmitted modes.}\label{f285TR12}
\end{figure}

\begin{figure}[htp]
\centering
\includegraphics[scale=.6]{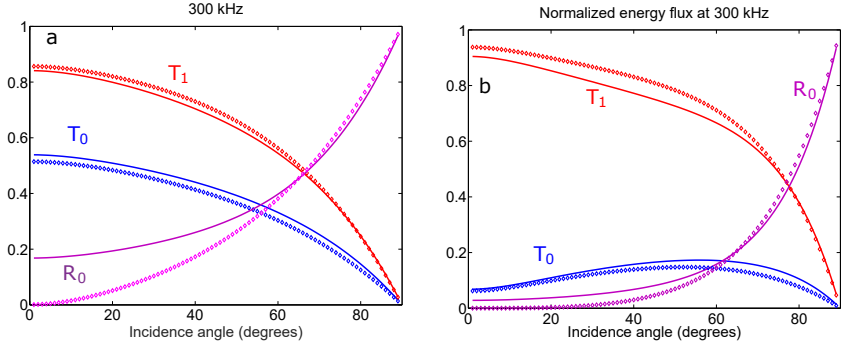}
\caption{Comparisons for frequency 300 kHz.}\label{f300TR}
\end{figure}

Fig. \ref{f285TR12} shows the calculated comparisons for frequency $f=285$ kHz, at which there is a switchover from one to two transmitted modes, at an angle of incidence around 30$^\circ$. The lower two of the plots show the comparisons in terms of energy flux. Good performance of the optimization scheme is again demonstrated, even close to the angle of incidence at which switchover occurs. It should perhaps be noted that the wave $m=1$ exists but is evanescent at smaller angles of incidence, with a rate of decay that approaches zero as the switchover angle is approached, corresponding to the boundary layer becoming arbitrarily thick. Our variational approximation is thus severely tested around this angle of incidence. Fig. \ref{f300TR} confirms good performance for $f=300$ kHz, at which frequency there are two propagating transmitted modes for all angles of incidence.

\begin{figure}[htp]
\centering
\includegraphics[scale=.6]{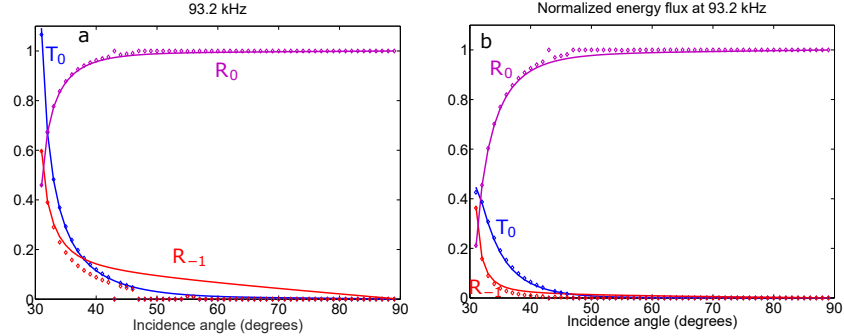}
\caption{Comparisons for incidence from a ``soft'' substrate, for which there is one propagating transmitted mode, negatively refracted, and two propagating reflected modes.}\label{f93TR}
\end{figure}

This section is concluded with some results for a different system, which was discussed by \cite{srivastava2016metamaterial}. The laminated material is the same but the aluminum half-space is replaced by one with shear modulus $\mu_0=0.4818$ GPa and density  $3000$ kg/m$^3$. Incidence from this half-space generates a single transmitted mode, which is refracted negatively for angles of incidence greater than $30^\circ$, and there are two propagating reflected modes. The relevant plots are shown in Fig. \ref{f93TR}.

\section{Conclusions}

In this paper we clarify the emergence and purpose of the evanescent waves which appear at the interface between a metamaterial and a homogeneous region in a model problem. We show that these evanescent waves form boundary layers on either side of the interface and that outside of these boundary layers the composite can be represented by appropriate infinite domain homogenized relations. We show that if one ignores the boundary layers then the displacement and stress fields are not continuous across the interface. Therefore, the scattering coefficients at such an interface cannot be determined through the conventional continuity conditions involving only propagating modes. We propose an approximate variational approach for sidestepping these boundary layers. The aim is to determine the scattering coefficients without the knowledge of the evanescent modes. Through various numerical examples we show that our technique gives very good estimates of the actual scattering coefficients, not only for the long wavelength region but far beyond it as well. The scattered energy is well estimated for all modes and at all angles of incidences - even in cases where multiple transmitted or reflected modes were present. The technique works well even in the case where negative refraction is occurring. 

\section{Acknowledgments}
A.S. acknowledges the support of the NSF CAREER grant $\#$1554033 to the Illinois Institute of Technology


%

\end{document}